\catcode`\@=11
\newif\if@fewtab\@fewtabtrue
{\count255=\time\divide\count255 by 60
\xdef\hourmin{\number\count255}
\multiply\count255 by-60\advance\count255 by\time
\xdef\hourmin{\hourmin:\ifnum\count255<10 0\fi\the\count255}}
\def\ps@draft{\let\@mkboth\@gobbletwo
    \def\@oddhead{}
    \def\@oddfoot
       {\hbox to 7 cm{$\scriptstyle Draft\ version:\ \draftdate$
       \hfil}\hskip -7cm\hfil\rm\thepage \hfil}
    \def\@evenhead{}\let\@evenfoot\@oddfoot}

\def\ceqno{\global\@fewtabfalse
    \ifcase\@eqcnt \def\@tempa{& & &}\or \def\@tempa{& &}
      \or \def\@tempa{&}
      \or\def\@tempa{}\fi\@tempa
{\rm(\theequation)}}

\def\aeqno#1{\global\@fewtabfalse
    \ifcase\@eqcnt \def\@tempa{& & &}\or \def\@tempa{& &}
      \or \def\@tempa{&}
      \or\def\@tempa{}\fi\@tempa
{\rm(\theequation,#1)}}

\def\label#1{\ifnum\draftcontrol=1
 \global\def\draftnote{$\scriptstyle #1$}\fi
 \@bsphack\if@filesw {\let\thepage\relax
   \def\protect{\noexpand\noexpand\noexpand}%
\xdef\@gtempa{\write\@auxout{\string
      \newlabel{#1}{{\@currentlabel}{\thepage}}}}}\@gtempa
   \if@nobreak \ifvmode\nobreak\fi\fi\fi
  \@esphack}

\def\alabel#1#2{\label{#1}\global\@fewtabfalse
    \ifcase\@eqcnt \def\@tempa{& & &}\or \def\@tempa{& &}
      \or \def\@tempa{&}
      \or\def\@tempa{}\fi\@tempa
{\hbox to 3cm{\phantom{\rm(\theequation,#2)}
\draftnote \hfil}\hskip -3cm {\rm(\theequation,#2)}}}

\def\clabel#1{\label{#1}\global\@fewtabfalse
    \ifcase\@eqcnt \def\@tempa{& & &}\or \def\@tempa{& &}
      \or \def\@tempa{&}
      \or\def\@tempa{}\fi\@tempa
{\hbox to 3cm{\phantom{\rm(\theequation)}
\draftnote \hfil}\hskip -3cm{\rm(\theequation)}}}

\def\eqnarray{\def\draftnote{{}}\global\@fewtabtrue
\stepcounter{equation}\let\@currentlabel=\theequation
\global\@eqnswtrue
\global\@eqcnt\z@\tabskip\@centering\let\\=\@eqncr
$$\halign to \displaywidth\bgroup\@eqnsel\hskip\@centering\@eqcnt\z@
  $\displaystyle\tabskip\z@{##}$&\global\@eqcnt\@ne
  \hskip 1\arraycolsep \hfil${##}$\hfil
  &\global\@eqcnt\tw@ \hskip 1\arraycolsep
$\displaystyle\tabskip\z@{##}$
\hfil  \tabskip\@centering&\global\@eqcnt\thr@@\llap{##}\tabskip\z@
\cr}

\def\endeqnarray{\@@eqncr\egroup
      \global\advance\c@equation\m@ne$$\global\@ignoretrue}

\def\@eqnnum{\hbox to 3cm{\phantom{\rm(\theequation)} \draftnote
                         \hfil}\hskip -3cm {\rm(\theequation)}}

\def\@@eqncr{\let\@tempa\relax
    \ifcase\@eqcnt \def\@tempa{& & &}\or \def\@tempa{& &}
      \or \def\@tempa{&}
      \or\def\@tempa{}
\fi\@tempa
\if@eqnsw
\if@fewtab\@eqnnum\fi
\stepcounter{equation}\fi\global
\@eqnswtrue\global\@eqcnt\z@\global\@fewtabtrue\cr}

\def\draftcite#1{\ifnum\draftcontrol=1#1\else{}\fi}

\def\@lbibitem[#1]#2{\item{}\hskip -3cm \hbox to 2cm
{\hfil$\scriptstyle\draftcite{#2}$}\hskip
1cm[\@biblabel{#1}]\if@filesw
     {\def\protect##1{\string ##1\space}\immediate
      \write\@auxout{\string\bibcite{#2}{#1}}}\fi\ignorespaces}

\def\@bibitem#1{\item\hskip -3cm \hbox to 2cm
{\hfil $\scriptstyle\draftcite{#1}$}\hskip 1cm
\if@filesw \immediate\write\@auxout
       {\string\bibcite{#1}{\the\value{\@listctr}}}\fi\ignorespaces}

\catcode `\@=11
\@addtoreset{equation}{section}
\def\theequation{\arabic{section}.\arabic{equation}}
\catcode `\@=12
\def\nsection#1{\section{#1}\setcounter{equation}{0}}

\def\ga{\gamma}         

\def\al{\alpha}

\def\la{\lambda}

\def\CA{{\cal A}}       \def\CB{{\cal B}}       
       \def\CE{{\cal E}}       
              
              \def\CL{{\cal L}}
\def\CM{{\cal M}}              \def\CO{{\cal O}}
\def\CP{{\cal P}}              
       \def\CT{{\cal T}}

\newcommand{\Nn}{{{\bf n}}}

\def\qq{ \begin{eqnarray} }
\def\qqq{ \end{eqnarray} }
\def\rr{ \qq }
\def\rrr{ \qqq }

\newcommand{\no}{\noindent}
\newcommand{\vs}{\vspace}

\newcommand{\hf}{{_1\over^2}}

\newcommand{\bea}{\begin{eqnarray*}}
\newcommand{\eea}{\end{eqnarray*}}
\newcommand{\ba}{\begin{eqnarray}}
\newcommand{\ea}{\end{eqnarray}}
\newcommand{\bee}{\begin{enumerate}}
\newcommand{\ene}{\end{enumerate}}

\def\ga{\gamma}         

\def\al{\alpha}

\def\la{\lambda}

\newcommand{\N}{\mathbb{N}}
\newcommand{\Z}{\mathbb{Z}}
\newcommand{\R}{\mathbb{R}}
\def\draftdate{\number\month/\number\day/\number\year\ \ \ \hourmin }

\global\def\draftcontrol{0}
\catcode`\@=12

\documentclass[12pt]{article}

\usepackage{mathbbol}
\usepackage{amssymb}

\renewcommand{\theequation}{\thesection.\arabic{equation}}
\def\theequation{{\thesection.\arabic{equation}}}

\begin{document}
\begin{center}

{\Large{\bf{Random walks in space time mixing environments}}}

\vs{ 0.5cm}

{\large{Jean Bricmont}}\footnote{Partially supported  by the Belgian IAP program P6/02.}

UCL, FYMA, chemin du Cyclotron 2,\\ 
B-1348  Louvain-la-Neuve, Belgium\\

\vs{0.2cm}

{\large{Antti Kupiainen}}\footnote{Partially supported  by the 
Academy of Finland.}

Department of Mathematics,
Helsinki University,\\
P.O. Box 4, 00014 Helsinki, Finland\\

\end{center}
\vs{ 0.2cm}

\begin{abstract}

We prove  that random walks in random environments,
that are exponentially mixing in space and time, are almost surely diffusive, in the sense that 
their scaling limit is given by the Wiener measure.
\end{abstract}

\date{ }


\vskip 1.3 cm

\nsection{The results}

Random walks in random environments are walks where the transition probabilities are themselves random variables (see  \cite{Sz, Z} for recent reviews of the literature).
The environments can be divided into two main classes: static and dynamical ones.
In the first case, the transition probabilities are given once and for all, and the walk can be ``trapped" for a long time in some regions because the transition probabilities happen to favour motion towards that region.
This may lead to anomalously slow diffusion in one dimension, as was shown by Sinai \cite{Si3}. 
In  \cite{bk, XX}, it is shown that, in three or more dimensions and for weak disorder (almost deterministic walks),  ordinary diffusion takes place.

In dynamical environments, the random transition probabilities change with time and trapping  does not occur, so that one expects ordinary diffusion to hold in all dimensions. Although simpler than the static environments, the dynamical ones are not trivial to analyze; see  \cite{DKL} for  recent  and general results and for references to earlier ones.

We consider in this paper a rather general class of space-time mixing
environments. This means that the transition probabilities at different times and
spatial points are weakly correlated and moreover the randomness
is weak. For such  environments we prove that the walks are diffusive,
almost surely in the environment measure. In particular
we do not assume  a Markovian structure of the environment.
We only assume that certain cumulants (or connected correlation functions) decay in a way that is typical of what happens in high temperature or weakly coupled Gibbs states.

Our motivation to study this class of models comes from the consideration of random walks in a deterministic, but ``chaotic" environment \cite{DL2}. As shown first by Bunimovich and Sinai, the invariant measures of suitably coupled hyperbolic dynamical systems correspond, via an extension of the SRB formalism,  to certain weakly coupled Gibbs states for a spin system on a space-time lattice \cite{BS, bk1, bk3, bk4, JM, JP, KL}. A walk whose transition probabilities are local functions of  such hyperbolic systems can be analyzed by the methods developed here.

Random walks in such deterministic environments emerge when considering
deterministic dynamics of a coupled map lattice with a global conserved
quantity ("energy"). The latter in turn can be viewed as a model of
coupled Hamiltonian systems where one would like
to prove diffusion and Fourier's law for heat transport. In
such models the environments will have more general correlations
than the Markovian ones and we expect to use the method
developed in this paper.
 This is discussed further at the end of this Section and in \cite{bk2}.

The method used in the proof consists in applying a Renormalization group scheme to iterate bounds, both on the size of the coupling between the  transition probabilities, and on  the size of their ``disorder", i.e. of their deviation from a deterministic walk. In the long time limit, the disorder tends to zero and the resulting deterministic walk behaves diffusively.

Turning to the precise models considered here,
let $\Omega^{\bf T}$
be the space of walks  $\omega=(\omega_0,\dots ,\omega_{\bf T})$, 
$\omega_t\in\mathbb{Z}^d$,  in time ${\bf T}$ and starting at $\omega_0=0$ and let 
the probability of a walk
 be defined as
\qq
P^{\bf T} (\omega) = \prod^{{\bf T}-1}_{t=0} p(t,\omega_t, \omega_{t+1}).
\label{1.2}
\qqq
The transition probabilities $p(t,u,v)$ of the walk 
 are taken to be random variables
 defined on some probability space $\Xi$, with distribution $\cal P$, satisfying the following assumptions:
  
  \vs{2mm}
  
  \noindent A.1. {\it Probability}.  $p(t,u,v)\geq 0$ and $\sum_v p(t,u,v)=1$.
  
  \vs{2mm}
  
 \noindent A.2. {\it Homogeneity and isotropy}  Let $\tau_s,\ \tau_w$
denote translations in time and space . We assume that $\tau_s \tau_w p$
has the same law as $p$. For  $R$ a rotation around the origin fixing
the lattice $\Z^d$ we assume that $p(t,u,v)$ and $p(t,u,u+R(v-u))$ are identically distributed  for all $t$, $u$, $v$.

  \vs{2mm}
  
 \noindent A.3.{\it Weak randomness}. Let $<->$ denote the expectation with respect to  $\cal P$ and define 
\qq
T(u-v)& :=&<p(t,u,v)>
\label{1.1}\\
b(t,u,v) &:=& p(t,u,v)- T(u-v).
\label{1.1.b}
\qqq
(where translation invariance was used).  Let, for $k\in \mathbb{T}^d$,
\qq
\hat T (k)= \sum_{u\in \mathbb{Z}^d} \exp(-iku) T(u),
\label{j12}
\qqq
be  the Fourier transform
of  $T$. 
We assume  that $\hat T$ is analytic in a complex neighborhood
of $ \mathbb{T}^d$ with
\qq
\hat T (k)=1-ck^2+\CO(|k|^4)
\label{smallk}
\qqq
in a neighborhood of origin where $c>0$ and
\qq
 |\hat T (k)|<1
 \label{less1}
 \qqq
  for $k\in \mathbb{T}^d\setminus 0$.

About  the "random" part
$b$, we will assume that it has  small   correlation functions decaying exponentially
in space and time as specified in eq. (\ref{1.4}) below.

  \vs{2mm}
  
 \no {\bf Remark}. Analyticity implies that $T(u)$ is exponentially decaying.
 Note that 
 for the transition matrix of nearest neighbour random walks,
$\hat T (k)={1\over d} \sum_{j=1}^d \cos k_j$,  which does not satisfy (\ref{less1})  at $k_j=\pi$, $\forall j$. 
However, if we take for $T$ the previous transition matrix composed with itself (i.e. nearest neighbour random walks after two steps), we get
$\hat T (k)=({1\over d} \sum_{j=1}^d \cos {k_j \over 2})^2$, and (\ref{less1}) holds (see \cite{bk}, Sect. 5, for a discussion of this point).
   
  \vs{2mm}
  
We now explain the assumptions made on the
random matrices $b$. We denote the pair
 $u,v$ by $z$ and $b(t,u,v)$ by $b(t,z)$. 
 Given $A\subset \mathbb{Z}$, introduce variables
 $z_t$ for $t\in A$ and define
\qq
b_A(z):= \prod_{t\in A} b(t,z_t).
\label{1.3}
\qqq

Since we need to deal with expectations of (\ref{1.3})
with possibly several copies of the same $b(t,z)$ we extend the definition (\ref{1.3})
to the disjoint union
\qq 
A=\coprod_{i=1}^mA_i
\label{1.3a}
\qqq
of $A_i\subset \N$, $i=1,\dots,m$.

 Recall the definition of  the connected correlation functions
 (or {\it cumulants})
\qq
 < b_A >^c \ = \sum_{\Pi \in \mathcal{P} (A)}(-1)^{|\Pi|+1} \prod_{B\in \Pi} < b_B >,
\label{1.31aa}
\qqq
where $\mathcal{P}(A) $ is the set of partitions of $A$.

We assume that these cumulants decay exponentially 
in the temporal and spatial  separations in  (\ref{1.31aa}). To spell this out
let,
for $B\subset \N$, 
$d(B)$ be the diameter of $B$, and for $A$ as in
(\ref{1.3a})  $d(A)=d(\cup A_i)$. 

For the spatial dependence,  let,
for a finite set $S\subset\R^d$,
$\tau(S)$ be the length of the shortest connected graph whose vertices  are
 a subset of $\mathbb{R}^d$ containing $S$. 
For $A$ and $z$ as above, define
\qq
\tau_A(z) :=  \sum_{t\in A}|u_t-v_t|   + \tau(S(z)),
\label{1.5}
\qqq
where $S(z)$ is the set of $u_t$ and $v_t$ in $z$
 (the reasons why we need this definition
for $S\subset\R^d$ instead of simply $S\subset\Z^d$ will be clear in the next section). 

 We assume that:
\qq
\|\langle b_A\rangle^c\|:=\sup_{z}
e^{\lambda \tau_A(z)}\mid < b_A(z) >^c \mid \leq \epsilon^{|A|}e^{-\lambda d(A)} ,
\label{1.4}
\qqq
for all $A$ of the form (\ref{1.3a}) with $m\leq n_0$, $\epsilon$ small enough and $\lambda$ large enough. Here, $|A|= \sum_{i=1}^m |A_i|$. 

  \vs{2mm}
  
We will study in this paper the large ${\bf T}$ properties
of the probability measure on paths defined by (\ref{1.2}). It will be convenient to realize them as measures
$\nu_{\bf T}$ on $C([0,1])$, the space of continous paths $\omega
:\;[0,1]\rightarrow {\mathbb R}^d$, by rescaling the time in
a standard way. Thus, given an $\omega\in \Omega$,
we obtain a piecewise linear path
\qq
{ \omega}(t)={\bf T}^{-{1\over 2}}(\omega_{i-1}
+({\bf T}t-i+1)(\omega_i-\omega_{i-1})), 
\label{j10}
\qqq
where $i-1={[{\bf T}t]}$ and $[{\;\;}]$ denotes the integral part.
$\nu_{\bf T}$ is the measure (\ref{1.2}),  transposed by (\ref{j10}),
on $C([0,1])$, and we will study the limit 
$
\lim_{{\bf T}\rightarrow\infty} \nu_{\bf T}, 
 $
also called the {\it scaling limit}, and its properties.
For reasons of convenience that will be explained in the next Section, we will consider below times of the form
${\bf T}=L^{2n}$ for $n\in {\N }$ and $L$ a fixed integer
 chosen later. We
will denote $\nu_{L^{2n}}$ by $\nu_n$ for short and expectations
in  $\nu_n$ by $\CE_n$. We let similarily $E_{\bf T}$
(or  $E_n$) refer to expectation in $P^{\bf T}$. They are related
simply by
\qq
\CE_nF(\omega(\cdot))=E_nF(L^{-n}\omega_{L^{2n}\cdot}),
\label{exps}
\qqq
for functions $F$ depending on $\omega$ restricted
to $L^{-2n}\mathbb Z$.

\vs{2mm}

We now state the main result concerning the scaling limit.  Let  $\nu^D $ be the Wiener measure with diffusion constant $D$
on paths $\omega\in C([0,1])$ with $\omega(0)=0$ and
$\CE^D$ be the corresponding expectation.  The scaling
limit of our walk is given by $\nu^D $ for almost all environments.
We prove that suitable
correlation functions converge, and this implies convergence of the diffusion
constant and of the finite dimensional
distributions (take $ f(x) = e^{ikx}$ below, and use Theorem 7.6 in \cite{Bil}).
 
\vs{2mm}
\noindent
{\bf Theorem.} {\it Let  $\cal P$
satisfy  A.1-A.3}.
{\it Then there is an  ${\epsilon}_0 > 0$  and  $\la _0$   such that, for
${\epsilon} < {\epsilon}_0$, $\la  > \la _0$  in (\ref{1.4}), there exists a  $D >
0\; such \; that$,} {\it for any any family $f_1\dots f_\kappa$, of polynomially bounded continuous functions, and} $t_1\dots t_\kappa\in [0,1],$
 $$ \lim_{n\rightarrow\infty}\CE_n\prod_i f_i(
\omega (t_i))=\CE^D\prod_i f_i(
\omega (t_i))$$
${\cal P}$- {\it almost surely}.

\vs{2mm}

\no {\bf Remark 1}. The diffusion constant
$D$  {\it satisfies}  (see (\ref{j4c}))
 \qq
 |D - D_0| \leq  C{\epsilon}^2,
 \label{j11}
\qqq
 where
 \qq
D_0=\sum_{u \in\mathbb{Z}^d}T(u)u^2.
\label{j16}
\qqq

\vs{2mm}

\no {\bf Remark 2}. With some extra work ${\cal P}$- almost sure
weak convergence also follows.
 \vs{2mm}

\no {\bf Remark 3}. Also with some more work, one should still be able to obtain the Theorem while replacing
 $ \tau(S(z))$
in the definition  (\ref{1.5}) of $\tau_A(z)$ by $\mbox{diam} (S(z))$. Indeed, the main point where the decay in $\tau_A(z)$ (see (\ref{1.4})) is used, is to control the integral (\ref{11}) below.
This should then allow an extension
of the result of  example 2 below to the coupled map lattices considered in \cite{bk3}, with smooth maps instead of analytical ones.

 \vs{2mm}

Let us finally give examples satisfying our assumptions.
 
 \vs{2mm}
 
 \no {\bf Example 1}. Let $\mu$ be the Gibbs measure
 for a high temperature Ising model on the space time
 lattice   $  \mathbb{Z}^{d+1}$ and let
 $s(t,x)$ be the spins. Let $p(s,x)$ be functions of $x\in   \mathbb{Z}^d$
and of the spins  $s(t,y)$ for $t$, $y$ close to $0$;  let the distribution induced by $\mu$ of
 $p(\cdot,x)$ be invariant under lattice rotations.
  Take
  $$
 p(t,u,v)=p(\tau_t\tau_u s,v-u)
 $$
 where $\tau_t$ and $\tau_u$ are translations in time and space.
 Then $p$ satisfies our assumptions. For a cluster expansion approach to  estimates like (\ref{1.4}), see
 e.g. \cite{Br, S, Mi}.
 
 This example 
 generalizes to $p$'s that are local and rotationally invariant functions of the variables distributed by completely analytic
 Gibbs states (see \cite{DS1,DS2,DS3,DS4} for definitions and examples of the latter). 
 
 \vs{2mm}

 \no {\bf Example 2}.  As an application of this extension to  completely analytic
 Gibbs states, one may consider, as in \cite{DL2},
  a {\it deterministic environment}
 generated by a chaotic dynamics. Let $\theta\in\CM=\mathbb{T}^{
 \mathbb{Z}^d}$ and let $f:\CM\to\CM$ be a {\it coupled analytic
 map}, as studied in \cite{bk1}. Let $\theta(t)=f^t(\theta)$, and
 $$
 p(t,u,v)=p(\tau_u\theta(t),v-u),
 $$
 where $p$ is {\it local} i.e depends on $\theta(t,x)$
 exponentially weakly in $|x|$. If $p$ is also analytic in $\theta$
 and if $\theta$ is distributed by the product of Lebesgue measures 
 on $\mathbb{T}^{
 \mathbb{Z}^d}$,
  then one can show, using the cluster expansion  in \cite{bk1}, that
   the assumption (\ref{1.4}) holds. This example
 will be discussed further in \cite{bk2}.

\nsection{The Renormalization group  }

The Renormalization group will allow us to replace the analysis of long time
properties of the walk by the study of a map, the Renormalization group map,
relating transition probability densities on successive
scales.

It will be convenient to extend the  transition probabilities $p(t,u,v)$
by constants to unit cubes centered at $u$ and $v$.  
Then the probability density to go from $u\in \R^d$ to $v\in \R^d$ in the time interval $I=[t,t']$ is given by
\qq
P_{[t,t']}(u,v,p) = \int d\omega_{t+1}\dots d\omega_{t'-1}
  \prod_{s=t}^{t'-1} p(s,\omega_s,\omega_{ s+1})
\label{1.7}
\qqq
with $\omega_t=u,\omega_{t'}=v$. We stressed in (\ref{1.7}) the dependence on the
random matrix $p$ and below we will use (\ref{1.7}) also for $p$'s that are not 
constant on unit cubes.

Let now $l\in \N$ and define a scaled transition probability density
\qq
R_lp(t,u,v) = l^dP_{[l^2t,l^2(t+1)]}(lu,lv,p)
\label{1.6}
\qqq
Then, if $l^2$ divides $t,t'$, by a simple change of variables,
\qq
P_{[t,t']}(u,v,p) =l^{-d} P_{[t/l^2,t'/l^2]}(l^{-1}u,l^{-1}v,R_lp).
\label{1.8}
\qqq

$R_lp$ are the {\it renormalized transition probability densities} at scale $l$. Note that they are constant
on $l^{-1}$ cubes centered at $(l^{-1}\Z)^d$. They are 
functions of $p$ and hence random matrices with a law
inherited from $p$. As $l\to \infty$ $R_lp$ controls the
long time behavior of the walk. For example, the
diffusion constant becomes
\qq
D(l^2)(p) = l^{-2}\int dy P_{[0,l^2]}(0,y,p)[y]_1^2
 = {\int}dy\ R_lp(0,0,y)[y]_l^2=D(1)(R_lp),
\label{1.17}
\qqq
where $[y]_l$ takes the value $x$ at the $l^{-1}$ cube
centered at $x\in (l^{-1}\Z)^d$.
Thus {\it the long time behavior is reduced to a time 1 problem for  $R_lp$, as
$l \rightarrow  \infty$}.  

$R_l$ is called the {\it renormalization group map}. Obviously
it is a semigroup, $R_{ll'}=R_{l}R_{l'}$ and the large $l$
limit is most conveniently studied iteratively. We choose
an integer $L>1$ and let $R:=R_L$ and $p_n=R^np$
i.e. $p_n=R_{L^n}p$.

To make a connection  to the scaling limit, let $F$ in
(\ref{exps}) depend on  $\omega$ restricted
to $L^{-2\ell}\mathbb Z$ and let $n=\ell+m$. Then, we get from
(\ref{exps}) 
$$
\CE_nF(\omega(\cdot))=E_n^p F(L^{-n}\omega_{L^{2n}\cdot}),
$$
where we  denoted the $p$ dependence explicitly, and then,
renormalizing by $l=L^m$,
\qq
\CE_nF(\omega(\cdot))=E_\ell^{p_m} F(L^{-\ell}\omega_{L^{2\ell}\cdot}).
\label{exps1}
\qqq
This relation will be used to prove the  Theorem.

 We will study the iteration
\qq
p_n\rightarrow p_{n+1}=Rp_n
\label{1.17a}
\qqq
where, from (\ref{1.7}, \ref{1.6}), we have 
\qq
Rp(t,u,v) =L^d \int d\omega_{I_t} \prod_{s\in I_t} p(s,\omega_s,
\omega_{ s+1})
\label{1.7b}
\qqq
with $I_t=[L^2t, L^2(t+1)-1]$,  $d\omega_{I_t}=d\omega_{L^2t+1}\dots
d\omega_{L^2(t+1)-1}$ and $\omega_{L^2t}=Lu,\omega_{L^2(t+1)}=Lv$.

\vs{2mm}

The map $R$ obviously preserves the properties A.1 and A.2, i.e.,
in particular, $\int dv p_n (t,u,v)=1$. As for A.3, let
us divide  $p_n$ into a ``deterministic" and a ``random" part
as in (\ref{1.1}) and  (\ref{1.1.b}): 
\qq
p_n(t,u,v) = T_n(u-v) + b_n(t,u,v) 
\label{1.18}
\qqq
where
\qq
T_n(u-v)  = \  < p_n(t,u,v) >
\label{1.19}
\qqq
 We have
${\int}dvT_n(v)= 1$ and   thus
\rr
{\int}dv \ b_n(t,u,v) =0 = \ < b_n(t,u,v) >. 
\label{1.22}
\rrr

The bulk of this paper consists in showing that  $b_n$ tends a.s. to zero as  $n
\rightarrow  \infty $,  whereas  $T_n$ tends to a Gaussian. The latter 
claim is  evident if $b = 0$.
Indeed, for a translation invariant $p$, the RG map (\ref{1.6}) is 
just a multiple convolution and becomes in terms of $\hat T$,  the Fourier transform (\ref{j12})
of  $T$,
\qq
 \hat T_{n+1}(k) =  \hat T_n({_k\over^L})^{L^2},
 \label{1.23}
\qqq
 i.e. 
\qq
\hat T_n(k) =\hat T({_k\over^{L^n}})^{L^{2n}}:=\hat\CT_n(k).
\label{1.24}
\qqq
By the assumption (\ref{smallk}) and (\ref{j16})
\qq
 \hat T(k)=1-(2d)^{-1}
{D_0 }k^2+\CO(|k|^4).
\label{j20}
\qqq
Hence, as $n\rightarrow\infty$, uniformly on compacts,
\qq
\hat T_n(k)
\rightarrow e^{- {_{D_0}\over^{ 2d}}k^2}
\equiv  \hat T^\ast_{D_0}(k) 
\label{1.26}
\qqq
where $T^\ast_{D}(x)$ is the unit time transition probability
density of the Wiener measure:
\qq
T^\ast_D(x) = (2\pi{ D/ d})^{-d/2} e^{- {_d\over^{2D}}x^2}  . 
\label{j15}
\qqq

Of course $b$ is not zero and, at each scale, $b_n$ will modify the
diffusion constant. Since $b_n$ goes to zero, we shall obtain
a sequence of approximations $D_n$, 
see (\ref{1.17}), to the true diffusion constant $D$.

  \vs{2mm}

The renormalization will allow us to iterate the following bounds
for $b_n$ and $T_n$. Let
\qq
\delta_n=L^{-n/2}e^{-\la}
\label{elln}
\qqq

  \vs{2mm}
\noindent {\bf Proposition 1.} {\it Under
the assumptions of the  Theorem, for all $A$ of the form (\ref{1.3a})
 \rr
 \|\langle b_{nA}\rangle^c\|\leq C \epsilon^{|A|}\delta_n^{d(A)}
\label{1.30}
\rrr
and moreover, for  $d(A)=0$,  we have
  \rr
 \sup_u\int dv e^{\hf\lambda \tau_A (z)} |\langle  b_{nA}(z)\rangle^c|
  \leq  C\epsilon^{|A|}\delta_n .
\label{1.30a}
\rrr
}

 \vs{2mm}
 
 As for the deterministic part, 
 we have
 \vs{2mm}
 
 \vs{2mm}

\noindent {\bf Proposition 2.}  {\it For $n\geq 1$, we have
  \qq
|T_n(x)|\leq Ce^{-|x|}, 
\label{j4b}
\qqq
moreover,
  \qq
|T_n(x)-T^\ast_D(x)|\leq C \delta_n e^{-|x|}, 
\label{j4a}
\qqq
where $D= \lim_{n\to \infty}\rho_n^2 D_0$, and
 \qq
|D-D_0|\leq C \epsilon^2. 
\label{j4c}
\qqq
 }
 
\vskip 0.2 cm

\noindent {\bf Remark on  the choice of constants.}
In the proofs, we use the letters $c$, $c'$ or $C$ to denote numerical constants independent
of $L$ (but that may depend on $\la$ and $n_0$) and $c(L)$ or $C(L)$ constants that do depend on $L$.
Those constants may vary from place to place, even in the same equation. Since $\la$ and $n_0$ are fixed
(and in fact, as we'll see in the proof of the  Theorem, $n_0$ could be taken equal to $2$), we will usually not indicate the dependence of constants on $\la$ or $n_0$.
We choose $L$ large enough so that we can always use $C \leq L$, or $C\leq L^\al$
for any given $C$ or $\al>0$ entering into our arguments. And we choose $\epsilon$ small enough so that
we can  use $C(L)\epsilon \leq 1$ 
for any $C(L)$. 

\nsection{Linearized RG}

From (\ref{1.7b}) and (\ref{1.18}), dropping the index $n$ and denoting $n+1$ by prime, we have the following recursion relation for $b_n$:
\rr
b' (t',u',v') = L^d  \int d\omega_{I_{t'}}  [\prod_{t\in I_{t'}} (T(\omega_{ t
}-\omega_{  t+1}) + b (t,\omega_{ t},\omega_{ t+1}))
-<->]
\label{1.32}
\rrr
In this section we will show how the bound in Proposition 1
iterates once the nonlinear relation (\ref{1.32})
is replaced by its linearization:
\rr
(\CL b)(t',u',v') = L^d 
\sum_{n}
  \int dudv
  T^n(Lu'-u)T^{L^2-n-1}(Lv',v)  b (t,u,v) 
\label{1.32a}
\rrr
(since $< b (t,u,v) >=0$, there is no subtraction as in (\ref{1.32})), where $t=L^2t'+n$ and $T^0(x)=\delta(x)$ (which takes values $L^{nd} $ on the  $L^{-n}$ cube centered at $0$, on scale $n$, since the transition probabilities are constant on cubes of side $L^{-n}$).

For  
each $t'\in A'$ pick $t(t')\in I_{t'}$ and define $n({t'})\in [0,L^2-1]$
by writing
$t(t')=L^2t'+n({t'})$.
Let $A$ be the collection of  $t(t')$ and
let $\Nn$ be the one of  $n({t'})$. The linearized RG is
then given by
 \qq
\langle(\CL b)_{A'}(z')\rangle^c  :=L^{d|A'|} \sum_{\Nn}
  \int dudv
  M_\Nn(u',u)N_\Nn(v',v) < b_A(z) >^c
\label{x1}
\qqq
where $z=(u,v)$, $z'=(u',v')$ and
\rr
 M_\Nn(u',u)=\prod_{t'}T^{n(t')}(Lu'_{t'}-u_t), \ \
N_\Nn(v',v)=\prod_{t'}T^{L^2-n(t')-1}(Lv'_{t'}-v_t)
\label{M}
\rrr
where $t=t(t')$ and the product runs over $t'\in A'$.

In this section, we first prove inductively the bound (\ref{1.30})  for the linearized part
of $b'$, i.e.:
 \rr
 \|\langle (\CL b)_{A'}\rangle^c\|\leq C \epsilon^{|A'|}\delta'^{d(A')}.
\label{w1}
\rrr

We need first
to express the exponent  $\tau_{A'}(z')$ in terms of 
$\tau_{A}(z)$.
Let $G_A$ be a connected graph with a set of vertices
including $S(z)$ 
and of length $\tau(S(z))$.
Let $E$ be the graph obtained by joining to
$G_A$ the lines with end points $Lu'_{t'}$ and
$u_{t}$ and $Lv'_{t'}$ and
$v_{t}$. Then its length is at least as large as $\tau(S(Lz'))=
L\tau(S(z'))$. Hence 
\rr
\tau(S(z'))\leq L^{-1}(\tau(S(z))+\sum_{t'}(|Lu'_{t'}-u_t|+
|Lv'_{t'}-v_t|)).
\label{x2000}
\rrr
Since also  $|u'_{t'}-v'_{t'}|\leq L^{-1}(|Lu'_{t'}-u_t|+
|Lv'_{t'}-v_t|+|u_t-v_t|)$ we obtain, using  (\ref{1.5}),
\rr
\tau_{A'}(z')\leq L^{-1}(\tau_A(z)+2\sum_{t'}(|Lu'_{t'}-u_t|+
|Lv'_{t'}-v_t|)).
\label{x2}
\rrr
Equations (\ref{x1}) and  (\ref{x2}) imply
\qq
I':= e^{\la\tau_{A'}(z')}|\langle(\CL b)_{A'}(z')\rangle^c|  &\leq&
 L^{d|A'|} \sum_{\Nn}
  \int du dv 
  \tilde M_\Nn(u',u)\tilde N_\Nn(v',v)
  \nonumber\\
  &&\cdot   
  e^{\la\tau_A(z)/L}|< b_A(z) >^c|
\label{x4}
\qqq
where $\tilde M$ and  $\tilde N$ are like $M$
and $N$ in (\ref{M}) but with $T^n$ replaced  by
\rr
\tilde T^n(u)=T^n(u)e^{{2\la \over L}|u|}.
\label{ttilde}
\rrr
Let first $d(A')>1$. Then, since $A$ contains one element in each $I_{t'}$, $t'\in A'$,
\rr
d(A)\geq L^2(d(A')-1)\geq \hf L^2d(A').
\label{dAbig}
\rrr
Since $A'$ contains at least $|A'|/n_0$ distinct times
we also have $d(A')\geq |A'|/n_0$. Thus we have
\rr
d(A)\geq  cL^2(|A'| +d(A'))
\label{damore}
\rrr
(with, say, $c= 1/(4n_0)$). To bound  $I'$, we use the inductive assumption 
 (\ref{1.30}) and 
 the $L^1$ bounds for $\tilde T$ in Lemma 2 (stated at the end of this Section).
The latter imply that the  $u$ and  the $v$ integrals are bounded by
$C^{|A'|}$. (\ref{damore}) implies 
$$
\delta^{d(A)}\leq \delta^{cL^2d(A')}e^{-c'L^2|A'|}.
$$ 
The sum over $\bf n$ is bounded by $L^{2|A'|}$; thus,
we obtain, for $L$ large enough, since, see (\ref{elln}), $\delta^{cL^2} \leq \delta'$, and $|A|=|A'|$,
\rr
I'\leq (CL^{2+d}e^{-c'L^2}\epsilon)^{|A'|} 
{\delta'}^{d({A'})}
 \leq \hf\epsilon^{|A'|}{\delta'}^{d({A'})}.
\label{x71}
\rrr

Let next  $d(A')=1$. This means that
\rr 
b'_{A'}(z')=\prod_{i=1}^kb'_{t'}(u'_i,v'_i)\prod_{j=1}^lb'_{t'+1}(u'_{k+j},v'_{k+j}),
\label{baprime}
\rrr
where both products have at most $n_0$ elements. 
Here, we need to use the property
 $\int dv b(t,u,v)=0$ to get the result. It allows us  replace
\rr
{T}^{L^2-n-1} (Lv^\prime-v)\to {T}^{L^2-n-1} (Lv^\prime -v)-{ T}^{L^2-n-1} (Lv^\prime -u).
\label{subs}
\rrr
in $N_\Nn$ for the terms with $L^2-n-1>0$. Let us assume, for the moment, that all $n(t')$ and all $L^2-n(t')-1$ are different from zero in
(\ref{M}). Since $\tau_A(z)\geq \sum_i|u_i-v_i|+\sum_j|u_{k+j}-v_{k+j}|$ we have
\rr 
e^{\la\tau_A(z)/L}\leq e^{\la\tau_A(z)}e^{-\sum_j|u_{k+j}-v_{k+j}|}.
\label{etau}
\rrr
Then, the right hand side of (\ref{x4}) is replaced by
\qq
I' \leq
 L^{d|A'|} \sum_{\Nn}
  \int du dv 
  \tilde M_\Nn(u',u) \tilde Q_\Nn(v',v,u)
  e^{\la\tau_A(z)/L}|< b_A(z) >^c|
\label{x4a}
\qqq
where
\rr
\tilde Q_\Nn(v',v,u)=\prod_{t'}S_{v_{t}-u_t}^{L^2-n(t')-1}(Lv'_{t'}-v_{t})
\label{x5}
\rrr
and
\rr
S_u^n(v)=|{T}^{n} (v)-{T}^{n}(v+u)|e^{2{\lambda \over L} |v|-c|u|/2},
\label{x6}
\rrr
where $c>0$ will be chosen below small enough (see (\ref{etaua})). Here, because of 
(\ref{etau}), we only need $c/2\leq1$.

Next, write $\tilde M_{\bf n}=\tilde M^1_{\bf n}\tilde M^2_{\bf n}$
corresponding to the two products in  (\ref{baprime})
and similarily for $\tilde Q_{\bf n}$. Using the pointwise bounds
of Lemma 2,
we bound
\rr 
\tilde M^1_{\bf n}\tilde Q^2_{\bf n} \leq  C^{|A'|}\prod_i (1+n_i)^{-d/2}\prod_j (1+L^2-n_{k+j})^{-(d+1)/2},
\label{M1}
\rrr
Using the $L^1$ bounds,
\rr 
\int dudv
\tilde M^2_{\bf n}\tilde Q^1_{\bf n}
\leq C^{|A'|}.
\label{M2}
\rrr
Thus
 \qq
I' \leq
 L^{d|A'|}(C\epsilon)^{|A'|} \sum_{\Nn}\delta^{d(A)}
 \prod_i (1+n_i)^{-d/2}\prod_j (1+L^2-n_{k+j})^{-(d+1)/2}.
 \label{x72}
\qqq

The sum can be controlled by the factor $\delta^{d(A)}$ since, for a given $d(A)$, there  are at most
$d(A)^{|A|}\leq d(A)^{2n_0}$ terms, since $|A|=|A'|\leq 2n_0$. If $d(A) \geq L^2/ 2$, 
  we bound the products in 
(\ref{x72}) by $1$; so, (\ref{x72}) is bounded
 by $(CL^d\epsilon)^{|A'|}(C\delta)^{L^2/2}$, and we use $|A'|\leq 2 n_0$, and $CL^{2dn_0}\delta^{L^2/2} \leq \delta'$, $d(A')=1$, to obtain (\ref{w1}).

If $d(A) < L^2/ 2$, since $d(A)\geq \max_i(L^2-n_i)+\max_j n_{k+j}$,
we have in the sum, $n_i \geq L^2/2$,  $L^2-n_{k+j} \geq L^2/2$, and 
the sum can still be  controlled by the factor $\delta^{d(A)}$. So,
the sum is bounded by $(C\delta)L^{-d|A'|}L^{-l}$ (since $d(A)\geq 1$) and 
\qq
I' \leq
(C \epsilon)^{|A'|} L^{-1}(C\delta)^{d(A')}\leq\hf \epsilon^{|A'|} \delta'^{d(A')},
 \label{x72a}
\qqq
using $C^{|A'|}L^{-\hf}\leq \hf $ (since $|A'|\leq 2n_0$), (\ref{elln}) and $d(A')=1$.

If some $n(t')$ or $L^2-n(t')-1$ equal zero in
(\ref{M}), then, since $d(A)\geq \max_i(L^2-n_i)+\max_j n_{k+j}$, we have $d(A)\geq L^2-1$.
We use $T^0(x)= \delta(x)\leq L^{nd}$ (on the $n$th scale). There are at most $2n_0$ such factors, and, by the definition (\ref{elln}) of $\delta_n$, $(CL)^{cn}\delta_n^{L^2-1} \leq \hf \delta'$ (with $c=2n_0d$),
for $L$ large enough.

This, (\ref{x72a}) and (\ref{x71}) prove (\ref{w1}) for $d(A')>0$.

Let finally $d(A')=0$. This means that we need to study
\qq
G_n(u,v)=\langle \prod_{i=1}^kb_{nt}(u_i,v_i)\rangle^c
\label{1}
\qqq
where the product has at most $n_0$ elements, and $k\geq 2$. 

Define the linear map 
\qq
{\cal L}_nG(u',v')=L^{kd}\sum_{t_1+t_2=L^2-1}(T_n^{\otimes k})^{t_1}G(T_n^{\otimes k})^{t_2}(Lu',Lv').
\label{5}
\qqq
${\cal L}_n$ is the part of the linearized RG
which involves $G_n$. 
The full RG is given by
\qq
G_{n+1}={\cal L}_nG_n+g_n+h_n
\label{3}
\qqq
where $g_n$ collects the 
the terms in the linear RG (\ref{x1}) 
with $d(A)>0$ and $h_n$ the
nonlinear contributions in eq. (\ref{1.32}).
The statement (\ref{1.30}) of Proposition 1, for $d(A')=0$, amounts to
showing
\rr
\|G_n\|\leq C\epsilon^k
\label{2a}
\rrr 
uniformly in $n$. Proceeding as above, we have
 \qq
\|g_n\|\leq C\epsilon^k\delta_n
\label{x1a}
\qqq
and in Section 4 we will prove that
\qq
\|h_n\|\leq C\epsilon^k\delta_n.
\label{x1aa}
\qqq
Thus, to prove (\ref{2a})
we need to control $\CL_n$.
Note that
$\CL_n$ is the derivative of the map 
\rr
G\rightarrow
L^{kd}G^{L^2}(L\cdot)
\label{6}
\rrr
computed at $G=T_n^{\otimes k}$.  Let  $\CL^*$ 
similarly be computed with $T_D^*$. The bound (\ref{j4a}) in Proposition
2 and (\ref{x2}) imply
\rr
\|\CL^*-\CL_n\|\leq C  \delta_n,
\label{7}
\rrr

Hence, to prove (\ref{2a}) it suffices
to bound $\|\CL^{*n}\|$ uniformly in $n$, as in (\ref{9}) below.
Indeed,  if this is the case, then,  (\ref{7}) implies
a uniform bound on   $\| \prod_{\ell=k}^n \CL_\ell \|$ in $k,n$, by $C'\prod_i (1+C\delta_i))$,
which is finite by (\ref{elln}).
Then, we get from  (\ref{x1a}),  (\ref{x1aa}), by iterating   (\ref{3}),  $$\| G_n\| \leq C \sum_{j= 0}^n   \| \prod_{\ell=j+1}^n \CL_\ell\| \epsilon^k \delta_j.$$ Since $\sum_j \delta_j <Ê\infty$ by (\ref{elln}), this implies (\ref{2a}).

Actually, $\|\CL^{*n}\|$ is not uniformly bounded, but, instead, we have the following Lemma, which allows us to conclude the proof of (\ref{2a}), since, by (\ref{1.22}), (\ref{8}) holds for $b_{nt} (u_i,v_i)$.

\vskip 2mm

\no{\bf Lemma 1.} {\it Let $G$ satisfy
\qq
\int dv_i G(u,v)=0
\label{8}
\qqq
for $i=1, \dots ,k$. Then, $\exists C<\infty$, such that
 \qq
\|\CL^{*n}G\|\leq C\|G\|
\label{9}
\qqq
uniformly in $n$. Moreover,
 \qq
\sup_u \int dv e^{\hf\la \tau_{A}(z)}|\CL^{*n}G(u,v)|
\leq CL^{-n}\log L^n \|G\|_1\leq \hf \delta_n \|G\|_1,
\label{9a}
\qqq
where $\|G\|_1$ denotes the norm in  (\ref{1.30a}).}
\vskip 2mm
 
Using this Lemma, we prove  (\ref{5}) following the proof of (\ref{2a}), using $\|G\|_1 \leq C \|G\|$, (\ref{x1a}), (\ref{x1aa}), and (\ref{9a}), which can be written as $\|\CL^{*n}G\|_1\leq \hf \delta_n \|G\|_1$.

\vskip 2mm
\no {\bf Proof of Lemma 1}.  Denote explicitly  the $L$ dependence of  $\CL^{*}_L$. 
We have  $\CL^{*n}_L=\CL^{*}_{L^n}$, because the map (\ref{6}) applied $n$ times is the same as 
(\ref{6}) applied once with $L$ replaced by $L^n$. Hence we need to study
the large $L$ behavior of $\CL^{*}_L$. 
The summand in (\ref{5}) is explicitly given by (dropping the
star)
\qq
L^{kd}\int dudv \prod_{i=1}^kT^{t_1}(Lu'_i-u_i)T^{t_2}(Lv'_i-v_i) 
G(u,v).
\label{10}
\qqq

Using (\ref{8}) we may again subtract $T^{t_2}(Lv'_i-u_i) $
from each 
 $T^{t_2}(Lv'_i-v_i)  $  when
$t_2>0$, which means that we replace $T^{t_2}(Lv'_i-v_i) $ in (\ref{10}) by
$T^{t_2}(Lv'_i-v_i)-T^{t_2}(Lv'_i-u_i) $. Recalling (\ref{1.5}), 
we write, instead of (\ref{etau}), 
$$
{\la\tau_A(z)\over L}\leq \la\tau_A(z)-\sum_i
 |u_i-v_i|-{\la \tau(S(z)\over 2},
$$
Since $\tau(S(z))$ is the length of a graph on $S(z)$,
$\tau(S(z))\geq |u_i-u_{i+1}|$ for all $i$, and thus $\tau(S(z))\geq \sum_i|u_i-u_{i+1}|/(k+l)$; so, combining the argument here with (\ref{x2}), we get: 
\rr
\la \tau_{A'}(z')\leq \la\tau_A(z)-c(\sum_i
 |u_i-v_i|+|u_i-u_{i+1}|)
+{2\la \over L}\sum_{t'}(|Lu'_{t'}-u_t|+
|Lv'_{t'}-v_t|)),
\label{etaua}
\rrr
where, since $k+l\leq 2 n_0$, $c$ depends only on $n_0$ and $\la$.

Then the supremum over $z'$ of  (\ref{10}) multiplied by $e^{\la\tau_{A'}(z')}$
is  bounded, using Lemma 2 (where we use bounds on $S_{v_{i}-u_i}^{t_2}(Lv'_{i}-v_{i})$, using definition (\ref{x6})), by 

\qq
L^{kd}(1+t_1)^{-kd/2}(1+t_2)^{-k(d+1)/2}
\int dudv e^{-c'(|Lu'_1-u_1|/\sqrt{t_1}
+|Lv'_1-v_1|/\sqrt{t_2})}e^{-c \sum_i (|u_i-v_{i}| +|u_i-u_{i+1}|)/2}
\|G\|.
\label{11}
\qqq
The factor $e^{-c \sum_i (|u_i-v_{i}| +|u_i-u_{i+1}|)/2}$ allows us to integrate over all the variables
(of which there are most $2n_0$), except one, say $u_1$.  And, using $|Lv'_1-v_1| + |v_1-u_1| \geq |Lv'_1 -u_1|$, 
for the integration over $u_1$, the integral is bounded  by:
\qq
C
\int du_1 e^{-c''(|Lu'_1-u_1|/\sqrt{t_1}
+|Lv'_1-u_1|/\sqrt{t_2} )}
\label{12a}
\qqq
which in turn is bounded by $C(1+t_i)^{d/2}$ where we use
$i=1$ if $t_1<L^2/2$ and $i=2$ if $t_1\geq L^2/2$.  Let us divide  the sum
over $t_1$ of  (\ref{11}) into one with $t_1<L^2/2$ and another with $t_1\geq L^2/2$.
In the first sum, we use $t_2\geq L^2/2$ to control the $L^{kd}$  factor, and in the second sum, we use
$t_1\geq L^2/2$.
The result is that the sum is bounded by:
\qq
C\sum_{0\leq t\leq L^2/2}( L^{-k}(1+t)^{-(k-1)d/2}+(1+t)^{-((k-1)d/2+k/2)})\|G\|.
\label{13}
\qqq
This is uniformly bounded in $L$ for all $d\geq 1$ and $k\geq 2$. The first claim follows. 

For the second one, we  integrate (\ref{10}) also
over $v'$, which absorbs the factor $L^{kd}$ through the change of variables $v' \to Lv'$. 
Using Lemma 2  for the $L^1$ norm of  $S_{v_{i}-u_i}^{t_2}(Lv'_{i}-v_{i})$ integrated over $Lv'_i$
and  (\ref{etaua}), we get that (\ref{10}), multiplied by $e^{\hf \la\tau_{A'}(z')}$, and integrated over $v'$,
is bounded by
\qq
C(1+t_2)^{-k/2}\|G\|_1
\int du \prod_{i=1}^k\tilde T^{t_1}(Lu'_i-u_i)e^{-c\sum|u_i-u_{i+1}|}
\label{14}
\qqq
Use Lemma 2  with $L^\infty$ norm for $k-1$ $\tilde T$'s
and $L^1$ norm for one $T$ to bound the integral by
$C(1+t_1)^{-(k-1)d/2}$. Altogether we end up with
a bound for the LHS of (\ref{9a}) (with $L^n$ replaced by $L$)
\qq
C\sum_{0\leq t_1\leq L^2}( 
(1+t_1)^{-(k-1)d/2}(1+t_2)^{-k/2)})\|G\|_1
\leq CL^{-1}\log L\|G\|_1.
\label{13a}
\qqq

\hfill $\Box$

\vskip 2mm

The proof of the following Lemma if deferred to Sect 4. 

 \vs{2mm}
 
 \no{\bf Lemma 2.} {\it Let $T= T_n$. There exists $C<\infty$, $c>0$ such that, for $L>L(\lambda)$, we have,
 using defintions  (\ref{ttilde}), (\ref{x6}),
$$
\tilde
T^m(u) \leq Cm^{-{d\over 2}}e^{-c|u|/\sqrt{m}},\ \
\|\tilde T^m\|_1\leq C,
$$
where $c$ can be chosen equal to $1$ for $n\geq 1$,
and
$$
\|S_u^m\|_\infty \leq C m^{-{d+1\over 2}},\ \   \|S_u^m\|_1\leq C m^{-\hf}
$$
for all $m\in [1,L^2]$. We also have $\|\tilde T^0\|_1\leq C$. }

\vs{2mm}

\nsection{Proof of Proposition 1}

As before, we drop the index $n$ and denote $n+1$ by prime.
Using the notation introduced in Section 1 we may expand the product over $t$, and 
 write (\ref{1.32}) as
 \qq
b' (t',z') = L^d \sum_{A}   \int dz  K_A({z'},z)  \Big(b_A(z) - < b_A(z)> \Big),
\label{1.33}
\qqq
where the sum runs over subsets of $I_{t'}$,
\qq
K_A({z'},z)= \prod^{l}_{i=0} T^{t_{i+1}-t_i-1} (v_i -u_{i+1}),
\label{1.34}
\qqq
with $T^0(u)=\delta (u)$.
$K$ depends on $z'$ through $v_0=Lu'$, $u_{l+1}=Lv'$. We have $|A|=l$.

Eq. (\ref{1.33}) leads to the following recursion relation for
the cumulants:

\vs{2mm}
\no {\bf Lemma 3.} {\it  Let $A'$ be of the form  (\ref{1.3a})
i.e. $A'= \coprod A'_i$. Then}
\qq
<b'_{A'} (z') >^c =L^{d|A'|} \sum_{\CA}
 \sum_{\Pi \in \mathcal{P}_{A'}^c(A)} 
  \int dz 
  \prod_{t'\in A'}    K_{A_{t'}}(z'_{t'},z)
   \prod_{B\in \Pi} < b_B(z) >^c.
\label{1.39}
\qqq
 {\it 
where $\CA=\{A_{t'}\}_{t'\in A'}$ is a family
of sets $A_{t'}\subset I_{t'}$ and $\CP_{A'}^c(A)$  is the set of partitions of  $A=\coprod A_{t'}$ that ``connect" $A'$
i.e. so that the following 
 graph is connected: its set of vertices is  $A'$ and its set of
edges are the pairs $\{t',t''\}$ such that, for some $B \in \Pi$, both
 $B \cap A_{t'}$ and $B \cap A_{t''}$ are nonempty.
}

\vs{2mm}

Now, the iteration of eq. (\ref{1.30}) follows the lines of Section 3, starting from
  (\ref{1.39}) instead of  (\ref{x1}). 
We need the analogues of (\ref{x2000}) and  (\ref{x2}). 
To state them we need some notation.

First, write, for $t' \in A'$, 
 $
{A_{t'}} =\{ t_{t'i}\ |\  i=1,\dots, |A_{t'}|  \}$.
Let
$z_{t_{t'i}}= (u_{t'i}, v_{t'i})$ and
$v_{t'0}=Lu'_{t'}$, $u_{t'|A_{t'}|+1} =Lv'_{t'}$.

It will also be important to single out the linear term
in  (\ref{1.33}). For this,   let 
$S'\subset A'$ consist of those $t'$ for which $A_{t'}$
consists of a single time, call it $t_{t'}$, and
let $S=\coprod_{t'\in S'}t_{t'}$. Note that Section 3 dealt with the case
where $S'= A'$.


 \vs{2mm}
\no {\bf Lemma 4.}  {\it Let $A' \setminus S' \neq \emptyset$.
For any value of 
$z$ in}   (\ref{1.39}), 
 \qq
\tau_{A'}(z') &\leq&
{1\over L}  (\sum_{B\in\Pi}\tau_B(z) + 2\sum_{t'\in A'} \sum^{|A_{t'}|}_{i=0} |v_{t'i}-u_{t'i+1}| )
\label{1.40}\\
d(S) &\leq&
\sum_{B\in\Pi}d(B) + 2 L^4|A'\setminus S'| 
\label{j50}\\
d(A') &\leq&\min\{ {2\over L^2} \sum_{B\in \Pi} d(B) + 
2 L^2|A'\setminus S'| , \  \sum_{B\in \Pi} d(B)\}.
\label{1.41}
\end{eqnarray}
 
 \vskip 2mm
 
Let 
 $\tilde K_{A_{t'}}({z'},z)$ be given by (\ref{1.34})
with
$T^t$ replaced by $\tilde T^t$ (see (\ref{ttilde})).
Then, inserting
(\ref{1.40}) into (\ref{1.39}),
we get
\qq
\|<b'_{A'}  >^c-<(\CL b)_{A'}  >^c\| \leq
L^{d|A'|} \sum_{S'\neq A' }  \sum_{S,\CB}
 \sum_{\Pi \in \mathcal{P}_{A'}^c(A)}
 I(S',S,\CB,\Pi)
  \label{j54}
\rrr
where the $\CB$ sum is over $A_{t'}$ with $t'\in A'\setminus S'$
i.e. such that $|A_{t'}|>1$.  We introduced also
\rr
I(S',S,\CB,\Pi)=\sup_{z'}
   \int dz
  \nonumber\\
 \prod_{t' \in A'}  \tilde K_{A_{t'}}({z'},z) 
   \prod_{B\in \Pi} e^{{\la\over L } \tau_B(z)}| < b_B(z) >^c|.
  \label{j541}
\qqq

To bound  (\ref{j541}), use again
$\sum_{t\in B}|u_{t}-v_t|\leq \tau_B(z)$, see (\ref{1.5}), which allows
us to replace each $\tilde K_{A_{t'}}$ by
\rr
 \tilde K_{A_{t'}}({z'},z)\prod_{t\in A_{t'}}e^{-c|u_{t}-v_t|}
\label{lastex1}
  \rrr
at the cost of replacing ${\la\over L}$ in the exponent
in (\ref{j541}) by $\la$.
The integral of (\ref{lastex1}) over $u$ and $v$ is
bounded by a convolution of $2|A_{t'}|$ $L^1$
functions whose $L^1$-norm  is $\CO(1)$, by Lemma 2. Thus, since $|A|=\sum_{t'} |A_{t'}|$,
\rr
I(S',S,\CB,\Pi)
 \leq C^{  |A|}\prod_{B\in\Pi} \|\langle b_B\rangle^c\|.
  \label{j54111}
\qqq
From our inductive assumption (\ref{1.30}), we get
\rr
\prod_{B\in\Pi} 
\|\langle b_B\rangle^c\|\leq (C\epsilon)^{|A|}\delta^{\sum d(B)} .
  \label{j541111}
\qqq
Recall that $\delta=L^{-\hf n}e^{-\la}$. Let first $n>0$. Taking convex combination of the bounds in
  (\ref{1.41}), we have
  $$
  d(A')\leq (1-x+2x/L^2)\sum d(B) + 2xL^2|A'\setminus S'|,
  $$
  and choosing $1-x+2x/L^2=(n-\hf)/(n+1)$,
  $$
  n\sum d(B)\geq (n+1)d(A')+\hf\sum d(B)-cL^2|A'\setminus S'|,
  $$
  where $c$ is independent of $n$, since $x=\CO(n^{-1})$, as $n \to \infty$.
 So, 
  \rr
\prod_{B\in\Pi} 
\|\langle b_B\rangle^c\|\leq C(L)^{|A'\setminus S'|}(C \epsilon)^{|A|}\delta'^{d(A')}\eta^{3\sum d(B)} .
  \label{j541112}
\qqq
where $\eta^3=L^{-1/4}$. For $n=0$ take $x=1$ and (\ref{j541112})
follows with $\eta^3=e^{-\la /4}$, using $e^{-\la L^2/4}\leq \delta'=\delta_1$.
  
Let us insert (\ref{j541112}), (\ref{j54111}) into (\ref{j54}), and  then turn to the four sums in eq. (\ref{j54}). To control them, we use the three factors 
$\eta^{\sum d(B)} $ in (\ref{j541112}).
For the  sum over partitions, we use the simple bound
\qq
\sum_{\Pi \in \mathcal{P}(A)} \prod_{B\in \Pi} \eta^{d(B)} \leq C^{|A|},
 \label{1.48}
\qqq
which holds for $\eta$ small enough,
since the left hand side of (\ref{1.48}) is bounded by
\qq
\prod_{t\in A}  (\sum_{t\in B \subset A} \eta^{d(B)}  ) \leq C^{|A|}.
 \label{1.49}
\qqq

Consider next the $\CB$ sum. 
Since each $A_{t'}\in\CB$
is a subset of size at least two of a set of $L^2$
points we have (recall that $|A|=|S'|+\sum_{t'\in A'\setminus S'}|A_{t'}|$)
\qq
\sum_\CB (C\epsilon)^{|A|}\leq (C\epsilon)^{|S'|}(C(L)\epsilon^2)^{|A'|-|S'|}.
 \label{j57}
\qqq

Finally, for  the sum over $S$, use (\ref{j50}) to write 
$\eta^{\sum d(B)}\leq (C(L))^{|A'|-|S'|}\eta^{d(S)}$.
Then, 
 write the elements
of $S$, $\{t_{t'}\}$, $t'\in S'$ as $t_1\leq t_2\leq \ldots \leq t_{|S'|}$
so that $d(S)=|t_{|S'|}-t_1|$.
Then
\begin{eqnarray}
\sum_S
\eta^{d(S)}\leq
 {\displaystyle\sum_{t_1\leq\ldots \leq t_{|S'|}}}
\eta^{|t_{|S'|}-t_1|} \leq  L^2C^{|S'|} \leq C(L)^{|A' \setminus  S'|} C^{|S'|},
\label{3.54}
\end{eqnarray}
since  at most $n_0$ times may coincide. The $L^2$ factor
comes from the sum over $t_1$ and the last inequality uses $A' \setminus  S' \neq \emptyset$. 

We need also to bound the factor $L^{d|A'|}$ in (\ref{j54}). We write $|A'|=|A' \setminus S'|+ |S'|$.
If  $d(S') \leq 1$, we have $|S'| \leq 2n_0$, and we can bound 
$L^{d| S'|}$ by  $C(L)^{|A' \setminus S'|}$, since $|A' \setminus S'| \neq 0$. If $d(S') >1$, we use  (\ref{dAbig}) to bound $|S'|\leq n_0 (d(S')+1) \leq c d(S)/L^2$, and use (\ref{j50}) for $d(S)$.  
Altogether, this gives, using the last factor $\eta^{\sum d(B)}$ in (\ref{j541112}),
\begin{eqnarray}
L^{(d+1)|A'|} \eta^{\sum d(B)} \leq C(L)^{|A' \setminus  S'|} L^{(d+1)|S'|}  \eta^{d(S)}\nonumber \\
\leq C(L)^{|A' \setminus S'|} L^{(d+1)|S'|}  \eta^{c'L^2 |S'|} \leq C(L)^{|A' \setminus  S'|} 
\label{3.54a}
\end{eqnarray}
So, we get:
\begin{eqnarray}
L^{d|A'|} \eta^{\sum d(B)}  \leq C(L)^{|A' \setminus  S'|} L^{-|A'|}.
\label{3.54ab}
\end{eqnarray}
where  the factor $L^{-|A'|}$ will be used now.

Combining    (\ref{j54111}), (\ref{j541112}),
(\ref{1.48}), (\ref{j57}), (\ref{3.54}) and  (\ref{3.54ab}), we get, for $\epsilon$ small,
\qq
 (\ref{j54})\leq  \delta'^{d(A')}
L^{-|A'|}\sum_{S' \neq A'}  (C\epsilon)^{|S'|}(C(L)\epsilon^2)^{|A'|-|S'|}\leq {_1\over^2}\delta'^{d(A')} \epsilon^{|A'|},
  \label{j54a}
\rrr
since the sum equals $(C\epsilon +C(L) \epsilon^2)^{|A'|}- (C\epsilon)^{|A'|} \leq (C'\epsilon)^{|A'|}$
and we use $L^{-|A'|}$ to control $C'^{|A'|}$.
Combining  (\ref{j54}), (\ref{j54a}),  and (\ref{w1}), (\ref{1.30}) is proven for $d(A')>0$. 

For $d(A')=0$, we obtain a bound  similar 
to (\ref{j54a}) on $h_{n+1}$ defined in (\ref{3}), with $\delta=\delta_n$ instead of
$\delta'^{d(A')}$, since all the terms in (\ref{j54}) have at least one power of $\delta$.
Using part of the factor $L^{-|A'|}$ in (\ref{j54a}), we can replace $\delta$ by $\delta'$
 which proves the bound  (\ref{x1aa}) for
$h_{n+1}$. Combining  (\ref{x1aa}) with  (\ref{x1a}) and  (\ref{9}) finishes the proof of  (\ref{2a}), i.e.
of  (\ref{1.30}) for $d(A')=0$, while using (\ref{x1aa}) with  (\ref{x1a})  and (\ref{9a}) finishes the proof of  (\ref{1.30a}).

\hfill $\Box$

We are left with the proofs of the Lemmas.

\no {\bf Proof of Lemma 3}. Using (\ref{1.33}) for $b'$, we get 
 \rr
<b'_{A'} (z') >  =  L^{d{|A'|} }\sum_{\{A_{t'}\}_{t'\in A'}} \int
  dz
  \prod_{t' \in A'}  K_{A_{t'}}(z'_{t'},z)\ \ < \prod_{t'}  \Big( b_{A_{t'}}(z) - < b_{A_{t'}}(z) >  \Big) >
\label{1.37}
\rrr
where $|A'|= \sum_i |A'_i|$ (note that here, up to  $n_0$ of the times $t'$ may coincide). To get connected
correlations, use first the inverse of  (\ref{1.31aa}):
\qq
< b_A(z) >  \ = \sum_{\Pi \in \mathcal{P} (A)} \prod_{B\in \Pi} < b_B(z) >^c.
\label{1.35}
\qqq
to obtain a recursion formula for
\qq
<\prod_{t'} ( b_{A_{t'}} (z)- < b_{A_{t'}} (z) >) > \ = 
\sum_{\Pi \in \mathcal{P}_{A'}(\coprod_{t'} A_{t'})} \prod_{B\in \Pi} < b_B (z) >^c
\label{1.36}
\qqq
where $\CP_{A'}$  is the set of partitions such that no $B\in \Pi$ is a subset of $A_{t'}$ for some $t'$.

Inserting (\ref{1.36}) into (\ref{1.37}) and denoting $A=\coprod_{t'} A_{t'}$,
we get:
\qq
<b'_{A'} (z') > =L^{d|A'|} \sum_{\{A_{t'}\}}
 \sum_{\Pi \in \mathcal{P}_{A'}(A)} 
  \int dz
  \prod_{t'\in A'}   K_{A_{t'}}(z'_{t'},z) 
   \prod_{B\in \Pi} < b_B(z) >^c.
\label{1.38}
\qqq
To prove (\ref{1.39}), consider a $\Pi$ in  (\ref{1.38}) and associate to it a  graph on $A'$
by connecting pairs $\{t',t''\}$ such that, for some $B \in \Pi$, both
 $B \cap A_{t'}$ and $B \cap A_{t''}$ are nonempty. Decompose that graph into connected components, $B'_i$, and write $A'=\cup_i B_i'$. This defines a partition of $A'$. Now, observe that the sum in (\ref{1.38}) factorizes over those connected components:
 $$
<b'_{A'} (z') >  =
\sum_{\Pi \in \mathcal{P} (A')} 
\prod_{B' \in \Pi} 
(L^{d|B'|} \sum_{\{A_{t'}\}_{t' \in B'}}
 \sum_{\Pi \in \mathcal{P}_{B'}^c(A)} 
  \int dz
  \prod_{t'\in B'}     K_{A_{t'}}(z'_{t'},z)
   \prod_{B\in \Pi} < b_B(z) >^c)
$$
where, for each factor in the product over $B'$, we write $A=\coprod_{t'\in B'} A_{t'}$.
Now, write (\ref{1.35}) with primes and observe that (\ref{1.35}) uniquely determines the connected
correlation function (because it is the inverse of (\ref{1.31aa})) to obtain (\ref{1.39}).\hfill $\Box$

\vs{2mm}

\noindent {\bf Proof of Lemma 4.}
Let $G_B$ be a connected graph whose set of vertices include $z_t$
for $t\in B$
and
whose length equals $\tau(S(z_B))$ (we denote the restriction of
$z$ to $B$ by $z_B$). Let $E$ be the graph
obtained by  joining to the union of the $G_B$ 
the lines with endpoints $v_{t'i}$ and $u_{t'i+1}$ for each $i=0,...,|A_{t'}|$, $t'\in A'$.
We claim that  $E$ is connected.

To see this observe first that any two points within the same $S(z_{A_{t'}})$ are connected by a path in $E$, since each $u_{t'i}$ is connected to $v_{t'i}$ (because they belong to the same $S(z_B)$), and each $v_{t'i}$ is connected to $u_{t'i+1}$ by the additional lines. 

Next, consider $w,\tilde w \in \cup_{t'\in A'} S(z_{A_{t'}})$. Since each $\Pi$ in (\ref{1.39}) connects $A'$, there exists a sequence $A_{t'_1},\ldots A_{t'_\ell}$ with $w\in S(z_{A_{t'_1}})$, $\tilde w \in S(z_{A_{t'_\ell}} )$, and a sequence $(B_i)^\ell_{i=1}$ such that ${B_i} \cap {A_{t'_i}} \neq \emptyset$, ${B_i} \cap {A_{t'_{i+1}}} \neq \emptyset$,
  $ i =1,\ldots,\ell-1$. So, we have
$S(z_{B_i}) \cap S(z_{A_{t'_{i+1}}}) \neq \emptyset$,
$S(z_{B_{i+1}} )\cap S(z_{A_{t'_{i+1}}}) \neq \emptyset$,
  $i =1,\ldots,\ell-1$.
 Since the graph $E$ connects each of the sets $S(z_{A_{t'}})$ and since there are points in $S(z_{B_i})$ and $S(z_{B_{i+1}})$ that belong to the same $S(z_{A_{t'}}) $, and thus, by the previous observation, are connected by a path in $E$, we see that there exists a connected path in $E$ joining  $w$ and $\tilde w$.  
 
Since the set of vertices of $E$ contains    $ S(Lz')$, and  $E$ is connected, its length 
is larger than $\tau(S(Lz') )=L\tau(S(z') )$.  By construction, the length of $E$ equals $\sum_{B\in\Pi}\tau(S(z_B)) +\sum_{t'\in A'} \sum^{|A_{t'}|}_{i=0} |v_{t'i}-u_{t'i+1}| $
so we get:
\rr
\tau(S(z'))   \leq {1 \over L} ( \sum_{B\in\Pi}\tau(S(z_B)) +\sum_{t'\in A'} \sum^{|A_{t'}|}_{i=0} |v_{t'i}-u_{t'i+1}|) .
 \label{b7}
\rrr
  Since also 
  $$|Lu'_{t'} -Lv'_{t'}|\leq \sum^{|A_{t'}|}_{i=0} (|u_{t'i}-v_{t'i}|+
|v_{t'i}-u_{t'i+1}|)$$    
the claim (\ref{1.40}) follows from the definition (\ref{1.5}).

Next we prove (\ref{j50}). 
 Let $\Pi_S\subset \Pi$ 
 be the set of  $B\in\Pi$ that contain elements of $S$. 
 Note that each $B\in \Pi_S$  has to contain elements of
 $A\setminus S$,
since $\Pi$ connects $A'$, unless  $S=A$, which
is not possible
 since $A'\setminus S'\neq \emptyset$
by assumption.
 
 Let then $A\setminus S \neq \emptyset$, so that we can assume that each
 $B\in \Pi_S$ contains elements of
 $A\setminus S$. For $B\in\Pi$, let $I_B=[s_B,t_B]$
 where $s_B$, (resp. $t_B$) is the minimal (maximal)
 time in $B$. Hence $d(B)=t_B-s_B$.
 Let $\sigma=\min_{B\in\Pi_S} s_B$, $\tau=\max_{B\in\Pi_S}  t_B$.
 Let $\bar I$ denote the smallest interval of
 $L^2\mathbb{N}$ containing $I$, an interval in $\mathbb{N}$.
Then the sets $B\in\Pi\setminus\Pi_S$
  connect
 the $\bar I_B$'s in $\Pi_S$. As a consequence
 $$
| [\sigma,\tau]\setminus\cup_{B\in \Pi_S}\bar I_B|\leq\sum_{B\in  \Pi\setminus\Pi_S}
d(B).
$$
Also, since $ |\bar I_B| \leq d(B) +2L^2$, $|\cup_{B\in \Pi_S}\bar I_B|\leq \sum_{B\in \Pi_S}d(B)+2L^2 |\Pi_S|$.
Since each  $B\in\Pi_S$ contains elements in $A\setminus S$,
$|\Pi_S|\leq |A\setminus S| \leq L^2|A'\setminus S'|$ and thus
$$
d(S)\leq |\tau-\sigma|\leq \sum_{B\in \Pi}d(B)+2 L^4|A'\setminus S'|.
$$
and (\ref{j50}) is proven. 
 
 Finally we 
 prove (\ref{1.41}).
 Since $S'\neq A'$,  
  then, as before,  each $B\in\Pi$ contains elements in $A\setminus S$ and hence $|\Pi|\leq L^2|A'\setminus S'|$.
For each $B$ we have $d(B)\geq |\bar I_B|-2L^2$.
 Also, $d(L^2A')\leq\sum_{B\in\Pi}  |\bar I_B|$. These imply
 $$L^2d(A')=d(L^2A')\leq\sum_{B\in\Pi} d(B)+2L^4|A'\setminus S'|$$
 which is our claim, since the  bound $d(A')\leq \sum_{B\in\Pi} d(B)$ holds trivially. 
 
  \hfill $\Box$

  \vs{2mm}

 
To prove  Lemma 2,   we need a Lemma on $T_n$, which will be proven in the next Section, since it will be also the basis of the proof of  Proposition 2. Note that the Fourier transform
 $\hat T_n$ is defined on the torus $\mathbb{T}_n=(L^n \mathbb{T})^d$.  The properties of  $\hat T_n$ are summarized in the following Lemma (where the domain of analyticity and the bounds are sufficient for our proofs but not optimal).

  
  \vs{2mm}

\noindent {\bf Lemma 5.}
{\it Let $\hat T$  be  as in} A.2.{\it   Then there exists
$r,c>0$ such that for all $n\geq 0$, $\hat \CT_n$, defined in (\ref{1.24}), 
is analytic in $|\Im k|< r^2L^{n\over 4}$   and,
for such $k$,
\qq
\hat \CT_n(k)=(1+\CO(L^{-2n}|k|^4))e^{-{D_0\over 2d}k^2}
\label{z1}
\qqq
if $|k|\leq rL^{n\over 4}$ and
\qq
| \hat \CT_n(k)|\leq e^{-cL^{n\over 2}}
\label{z2}
\qqq
otherwise. 

Moreover, under
the assumptions of the  Theorem,  $T_n$ 
can be
expressed as
\qq
\hat T_n(k) = \hat {\cal T}_n (\rho_n  k)+ \hat  t _n (k)
 \label{j2}
 \qqq
\noindent
where
\qq
|\rho_n - 1| \leq  C {\epsilon}^2,
 \label{j4}
  \qqq
  for $C<\infty$, with $\rho_0=1$.
  We have $t_0=0$,
\qq
  |\hat t _n(k)| \leq \epsilon 
\delta_n   |k|^4,
\;\;\; {\rm for} \;\; |k|\leq 1.
  \label{j5}
  \qqq
  and
  \qq
 |t_n (x)|  \leq \epsilon  \delta_n  e^{-2|x|}.
 \label{j7}
\qqq
 }

 \vs{2mm}
\noindent {\bf Proof of Lemma 2.}  
Let $n\geq 0$ and $\CT:=\CT_n$. Let $m\in [1,L^2]$
and consider $\CT^m(x)=
\int dk \hat \CT(k)^me^{ikx}$. Shift the $k$ integration
by $ip$ (to be precise, by $\pm ip$ for each coordinate $j$, depending on the sign of $x_j$), with $p^2=a^2/m<r^4$.  Then we get, using
 (\ref{z1}) and  (\ref{z2}),
$$\CT^m(x)\leq Ce^{-a|x|/\sqrt{m}}(e^{ca^2}m^{-{d\over 2}}+
e^{-cm}),$$
where we divided the integral into $|k| \leq 1$ and $|k|>1$, and used
$|e^{-{D_0\over 2d}k^2}| \leq e^{-c|k|^2}$ for $|k|>1$, $|\Im k| < r^2$.
  Clearly, for $L$ large, we may, for all $m\in [1,L^2]$,  choose $a=a(m)=\min(r^2\sqrt{m}/2, 3\lambda)$,
 so that, for all $1\leq m \leq L^2$ and $L$ large enough,  $a/\sqrt{m}-2\la/L\geq c>0$ and $a^2/m<r^4$. Hence,
  \rr
  \tilde\CT^m(x)\leq Cm^{-{d\over 2}}e^{-c|x|/\sqrt{m}}.
  \label{z3}
  \rrr
  where  $\tilde\CT$ is defined as in (\ref{ttilde}) and $C$ and $c$ depend only on $T$ and $\la$.
  
  From  (\ref{j7}) we obtain that $\hat t(k)$ is analytic  for $|\Im k|<2$ and is bounded by
  $C \epsilon \delta_n$. Since this is less than $m^{-{d\over 2}}$ for $m\leq L^2$,
 we can repeat for  $\hat T(k)$    the  argument given for $\hat \CT(k)$. For $n\geq 1$, the domain of analyticity of $\hat \CT(k)$ can be taken as large as one wants (by choosing $L$ large)
and  we can choose $a$ large enough so that we can have $c=1$ in (\ref{z3}). Since $c=1$ is $L$ independent, the constants $C$ in (\ref{z3})  still depends  only on $T$ and $\la$.
    This proves the  first two claims of Lemma 2.
  
 For the other two claims, observe that, if we prove them for $|u|=1$, then we can interpolate between $v$ and $v+u$ by steps of size $1$ and obtain the result. The exponential factor in (\ref{x6})  $e^{-c|u|/2}$ controls both the $e^{2\la|u|/L}$ coming from the interpolation (for $L$ large) and the number $|u|$ of interpolation steps.
 Now, for $|u|=1$,
 we shift again the integration contour: 
    $$
    |{\cal T}^m(v+u)-{\cal T}^m(v)|\leq e^{-p|v|}
    \int \hat \CT(k)^m |e^{(p-ik)u}-1| dk,
    $$
   (where, with an abuse of notation, $p$ denotes a number in $p|v|$ and a  vector in $pu$).
    We take $p=a/\sqrt{m}$, with $a=a(m)$ as above,
    Using  (\ref{z1}) and  (\ref{z2}) we bound
    the integral by
     $$
       \int_{|k|\leq 1} e^{-cmk^2}
      |e^{(p-ik)u}-1| dk+Ce^{-cm}
    $$
The integral is bounded by  $C(p+m^{-1/2})m^{-d/2}$. Altogether we get, since $a\leq 3 \lambda$,
  $$
    |{\cal T}^m(v+u)-{\cal T}^m(v)|\leq C(p+m^{-1/2})m^{-d/2}e^{-p|v|} \leq Cm^{-(d+1)/2}e^{-a|v|/\sqrt{m}},
    $$   
  where again $C$ depends only on $T$ and $\la$. We  then  get the estimates
   for $\tilde T$ as above, since
   $\hat t$ gives  corrections of order $\epsilon \delta_n$.

\nsection{Proof of Proposition 2}

We start with the 
 
 \vs{2mm}
\noindent {\bf Proof of Lemma 5.}  Our assumptions on $T$
imply that for $r$ small enough, $|\hat T(k)|\leq \rho(r)<1$ for $|\Re k|>r$,
$|\Im k|\leq r^2$; this implies (\ref{z2}) for $|k|> r L^n$, $|\Im k| \leq r^2 L^{n \over 4} $. Write (\ref{j20}) as $\hat T(k)=e^{-{D_0\over 2d}k^2} (1+ \CO(|k|^4))$
for $|k| \leq r$. This implies (\ref{z1}) for $|k| \leq rL^n$, in particular for $|k| \leq rL^{n \over 4}$, $|\Im k| \leq r^2 L^{n \over 4} $.
Since $|e^{-{D_0\over 2d}k^2}|= e^{-{D_0\over 2d}((\Re k)^2-(\Im k)^2)} \leq e^{-{D_0\over 4d}|k|^2}$
for $\hf |k| > |\Im k|$, the claim (\ref{z2}) holds also for  $rL^{n \over 4} <|k| \leq  r L^n$, $|\Im k| \leq r^2 L^{n \over 4}, if $ if  $ r$
is taken small enough.

For the other statements, write again $T$ for $T_n$ and $T'$ for $T_{n+1}$.  
We have from (\ref{1.33})
 \qq
 T'(x)=L^dT^{L^2}(Lx)+\beta(x)
\label{51}
\qqq
with
 \qq
\beta(x) = L^d \sum_{A}   \int dz  K_A({z'},z)  < b_A(z)> ,
\label{52}
\qqq
where $z'=(0,x)$ and only $A$'s containing  distinct times enter. 
Note that  (\ref{52}) collects the averages that have been subtracted in   (\ref{1.33}) and therefore 
enter the definition  of $T'$. Now use (\ref{1.35}); since $A$ contains only distinct times, 
 each $B$ in  (\ref{1.35}) has $d(B)>0$.   Using (\ref{1.40}) and $z'=(0,x)$ to bound $e^{2 |x|}$,
 Proposition 1 and Lemma 2  immediately imply the bound
\qq
|\beta(x)|\leq C(L)\epsilon^2\delta e^{-2 |x|} \leq \hf \epsilon\delta' e^{-2 |x|},
\label{53}
\qqq
where $\epsilon^2$ comes because  $|A|\geq 2$ and we use $C(L)\epsilon \leq 1$.

In terms of Fourier transform, (\ref{51}) reads:
\qq
 \hat T'(k)=\hat T(k/L)^{L^2}+\hat\beta(k)
\label{54}
\qqq
By (\ref{53}),  $\hat \beta$ is analytic
in $|\Im k|\leq 3/2$ and bounded there by $C(L)\epsilon^2\delta$.
By the isotropy assumption A.2 in Section 1, the Taylor expansion
reads
\rr
\hat\beta(k)=\zeta k^2 +\CO(|k|^4).
 \label{54a}  
 \rrr
 We used $\hat\beta(0)=0$ which follows from (\ref{54})
 and $\hat T(0)=1=\hat T'(0)$.
 By Cauchy's theorem
\rr
|\zeta |\leq C(L)\epsilon^2\delta,
   \label{j26}
   \rrr
and 
\rr
|\hat \beta(k)-\zeta k^2 | \leq  C(L)\epsilon^2\delta |k|^4,
   \label{j27}  
\rrr
for  $|k| \leq 1$.

The $\beta$ term  will ``renormalize" the effective
diffusion constant  $D= \rho D_0$.
We set
\qq
\rho'^2 = \rho^2- {2d\zeta  D_0^{-1}}.
   \label{j28}
    \qqq
\noindent
 (\ref{j26}) and (\ref{elln}), which implies the convergence of $\sum_n \delta_n$, then imply the bound (\ref{j4}).
 
 Consider next the first term in (\ref{54}) (recall (\ref{j2})):
\rr
\hat{T}({_k \over^L})^
{L^2}=(\hat{\cal T}(\rho{_k \over^L})+
\hat{t} ({_k \over^L}))^{L^2}
:= \hat{\cal T}(\rho{_k \over^L})^{L^2}
+\hat{\tau}({_k \over^L}).
\label{j34}
\rrr
Since  $\hat T'(k)= \hat{\cal T}'(\rho'k)+t'(k)$
 we get from  (\ref{54})  
and  (\ref{j34})  that
\rr
\hat{t}' (k)=  {\hat{ \tau}}({_k \over^L}) + \hat\beta(k)+ \hat{r}(k),
\label{j32}
\rrr
where
\rr
{\hat r}(k) := \hat{\cal T}'(\rho k)- \hat{\cal T}'(\rho'k),
   \label{j30}
    \rrr
since, by definition (\ref{1.23}), $ \hat{\cal T}'(\rho k)=  \hat{\cal T}(\rho{ k \over L})^{L^2}$.   

We need to show that $t^\prime$ satisfies (\ref{j5}) and (\ref{j7}) with $\delta'$.   Consider (\ref{j5}) first. From (\ref{j34}), we have
\qq
\hat{\tau} ({_k \over^L}) = \sum^{L^2}_{m=1} ( ^{L^2}_m ) \hat{t}  ({_k \over^L})^m  \hat{{\cal T}}  (\rho{_k \over^L})^{L^2-m}
 \label{001}
 \qqq
By (\ref{j5}), $| \hat{t} ({k\over^L})| \leq \epsilon \delta L^{-4} {|k|^4}$ 
and by Lemma 5, $|\hat{\cal T}|$ is bounded  for $|k| \leq 1$.
 Hence 
\rr
| \hat{\tau} ({_k\over^L})| \leq 
C\epsilon \delta L^{-4} {|k|^4}(L^2+C(L)\epsilon \delta)
\leq \hf  \epsilon \delta' |k|^4,
\label{002}
\rrr
 for $|k| \leq 1$. This bounds the first term in (\ref{j32}).

Using (\ref{z1}, \ref{z2}) to bound the derivative of
  $\hat{\CT}'$,  (\ref{j28}) and (\ref{j26}) imply
\rr
|\hat{r}(k)| \leq C(L)\epsilon^2\delta
 \label{j37}
\rrr
for $|k|\leq 2$ (note that we apply (\ref{z1}, \ref{z2}) to $n\geq 1$ here, i.e. we can assume that $r^2 L^{1/4}$ is large enough). Note that $\hat{r}(k)$   satisfies
$\hat{r}(k)= -\zeta k^2 + \CO(k^4)$ so that 
we infer from (\ref{54a}) $\hat\beta (k) + \hat{r}(k)=\CO(|k|^4)$. 
Combining this with   (\ref{j27}), (\ref{j37})
and a Cauchy estimate yields
\rr
| \hat{\beta} (k)+\hat{r} (k)| \leq 
C(L)\epsilon^2\delta|k|^4 \leq \hf \epsilon \delta' |k|^4,
\label{003}
\rrr
 for $|k| \leq 1$.  Then, (\ref{002}) and  (\ref{003}) imply  (\ref{j5}).

Next, we prove (\ref{j7}).  Combining (\ref{z1}, \ref{z2}) with $n\geq 1$ and (\ref{j28}, \ref{j26})
 with (\ref{j30}), we infer
\rr
|{r}(x)| \leq C(L) \epsilon^2\delta e^{-2|x|} \leq {1 \over 4} \epsilon \delta' e^{-2|x|}.
 \label{j37a}
\rrr
As for $\tau$, we have from (\ref{001}),
$$\tau(x)=L^d\sum_{m=1}^{L^2} ( ^{L^2}_m)
 ( \CT^{L^2-m}(\cdot /\rho)t^m)(Lx).$$
 Consider first the $m=1$ term. Its Fourier transform is given by the $m=1$ term in 
 (\ref{001}). By shifting the integration contour the $m=1$ is
thus bounded by
 $$
L^2e^{-2|x|}\int | \hat{t}  ({_k \over^L})  \hat{{\cal T}}  (\rho{_k \over^L})^{L^2-1}|dk,
 $$
 where $|\Im k|=2$.
 Use (\ref{j5}) for $|k| \leq L$, and (\ref{z1}), (\ref{z2}) (for $n\geq 1$, since $t_0=0$)  to bound the integral over $|k| \leq L$ by $C \epsilon \delta L^{-4}  $. For $|k| \geq L$, we use the fact that, by  (\ref{j7}),  $|\hat{t}  ({_k \over^L})|$
 is bounded by $C \epsilon \delta$, and that, by (\ref{z1}), (\ref{z2}), the integral of $ \hat{{\cal T}}  (\rho{_k \over^L})^{L^2-1}$ over $|k| \geq L$ is less than $C \exp(-c L^\hf)$. Hence altogether the $m=1$ term is bounded by
$$
C(L^{-2}+e^{-cL^\hf})\epsilon\delta e^{-2|x|}.
$$
 The $m \geq 2$ terms in (\ref{j37a}) are easily bounded,  using (\ref{z1}, \ref{z2})
and (\ref{j7}) and only add   $\delta$ to $(L^{-2}+e^{-cL^\hf})$.
Hence $\tau (x)$ is bounded by the right hand side of (\ref{j37a}). 
Combining these bounds with   (\ref{53}),
eq. (\ref{j7}) follows for $t'$.
\hfill $\Box$

\vs{2mm}

Now, the proof of Proposition 2 is straightforward:

\vs{2mm}

\noindent {\bf Proof of Proposition 2.}  
We get  (\ref{j4b}) by combining (\ref{z1}, \ref{z2}) for $n\geq 1$ and (\ref{j7}).
To show (\ref{j4a}), we write, using the definition  of $D=\lim_{n\to \infty} \rho^2_n D_0$,
$$
T_n(x)-T^\ast_D(x)=  \CT_n(\frac{x}{\rho_n})-T^\ast_D(x) + t_n (x).
$$
From (\ref{j28}), (\ref{j26}), we get that (for $\epsilon$ small) $|\rho^2_n D_0-D| \leq \delta_n$.
Then, we use (\ref{z1}, \ref{z2})  and bound the derivative of $\CT_n (k)$ to get (\ref{j4a}) for the first term. We use
(\ref{j7}) for the second. Finally, we get
(\ref{j4c})  from (\ref{j4}).
\hfill $\Box$


\nsection{Proof of the Theorem}

Since the functions $f_{i}$ and the paths $\omega$ are continuous, it is enough to prove
the Theorem  for any given family ${\bf f} =(f_i)_{i=1}^\kappa$ and for all sets of times ${\bf  t}= (t_i)_{i=1}^\kappa$, where $t_{i}$ belongs to the dense set $ \cup_\ell L^{-2\ell} {\N}$ . So, let 
$t_{i} \in L^{-2\ell} {\N}$, $i=1, \dots, \kappa$, and $\ell$ fixed. We use (\ref{exps1}) and $p_m= T_m + b_m$ as in
(\ref{1.33}) to get (recall that $n=\ell+m$)
 \rr
 \CE_n\prod_i f_i(
\omega (t_i))=
  \sum_{A}   \int 
  dz dv' K_{mA}({z'},z)  b_{mA}(z) 
 \prod^\kappa_{i=1} f_i (L^{-\ell} x_i):=\sum_A I_n(A,{\bf f, t},\ell)
 \label{y1}
\qqq
 where notation is as in (\ref{1.34})
 with $L$ replaced by $L^\ell$, $v_0=0$ and $v_{l+1}=v'$.
 The $x_i$'s form a subset of the $u_j,v_j$'s. It is useful to remember that the product over $T$ and $b$ in (\ref{y1}) is ordered over the time interval $[0, L^{2\ell}]$.

 The terms $I_n(A,{\bf f, t},\ell)$  are random
 variables.
  We show first that,  for any ${\bf f}$, there is a set $\CB$ of measure one
  such that, for $b\in\CB$, $\lim I_n(A,{\bf f, t},\ell)=0$, 
  for all $A\neq\emptyset$ and all ${\bf  t}$.

First note that, if
 $f$ is polynomially bounded, then, for all $\ga>0$, we
 can find a constant $C(\gamma,f)$ such that
$| f(x) | \leq C(\ga,f) \exp({\ga |x|})
$. Thus writing $f_i=C(\ga,f_i)f_i'$ it is enough to prove
the claim for $f_i$ such that $C=1$; $\gamma$ will be
chosen below.

Next, since $v_0=0$ and $x_i$
is one of the $u_j,v_j$,
\qq
|x_i|\leq \sum_{j=1}^\ell (|v_{j-1} - u_{j}| +  |u_j - v_{j}|).
\label{y2}
\qqq
Therefore, writing $I_n(A,{\bf f, t},\ell)=I_n(A)$,
\rr
 \langle I_n(A)^2\rangle \leq
 \int dz dv' 
\tilde K_{m\CA}({z'},z) |\langle \tilde b_{m\CA}(z)
 \rangle |
   \label{y3}
\qqq
where, in $\tilde K_m$, $T_m(u)$ is replaced by $e^{\kappa \gamma |u|}T_m(u)$
and $b_m(t,u,v)$ by $e^{\kappa \gamma  |u-v|}b_m(t,u,v)$
and  $\CA= A \coprod A $ (note that there are 
twice
as many variables $z$ and $v'$ compared to  (\ref{y1})).

Next, expand the expectation value in (\ref{y3}) in terms of connected correlation functions, using (\ref{1.35}). We need
to bound then
\rr
J:= \int dz dv' 
\tilde K_{m\CA}({z'},z)\prod_{B\in\Pi} |\langle \tilde b_{mB}(z)
 \rangle^c |
   \label{y4}
\qqq
where $\Pi\in\CP(\CA)$, since the number of terms in (\ref{1.35}) depends on $|A|$, i.e. on $L^{2\ell}$. By (\ref{1.5}),  
$ \sum _{ t_i \in B} |u_i - v_{i}| \leq \tau_B (z) $. Thus
for $\kappa \gamma <\la/2$,  
\rr
J\leq \int dz dv' 
\tilde K_{m\CA}({z'},z)\prod_{B\in\Pi}e^{\hf \la\tau_B(z)} |\langle  b_{mB}(z)
 \rangle^c |.
   \label{y5}
\qqq
Note that each time appears in $\CA$ at most twice. 
If there are  factors with $d(B)\neq 0$, we use 
 (\ref{1.30}) and the remaining integrals consist of order $\ell$
 convolutions of $\tilde T$, each of which is bounded by $\|\tilde T\|_1\leq C$,  by Lemma 2,
 for $\kappa \gamma <2\la/L$ (see  (\ref{ttilde})). If only $B$'s with $d(B)=0$ occur, use $\|\tilde T\|_1\leq C$
 for all the factors $ \tilde T$ occuring after the last $B$ and use  (\ref{1.30a}) for that one (we necessarily have an integral over $v$ here, since we integrate over the last $v$ variable, denoted $v'$ in  (\ref{y1})).
 The result is:
  \rr
 \langle I_n(A)^2\rangle \leq
 C(\ell,L)\epsilon^{2|A|}\delta_m.
    \label{y6}
\qqq

By Chebyshef's inequality we get
\qq
P(|I_n (A)|>1/k) \leq C(\ell,L) k^2 \epsilon^{2|A|}\delta_m.
\label{008}
\qqq
Since, by (\ref{elln}), $\sum_m \delta_m < \infty$, we get, by the first Borel-Cantelli lemma, that, for any given
${\bf f,t}$, $A\neq\emptyset$ and $k\in\mathbb N$, 
there is a set of measure one, $\CB_k({\bf f,t},  A)$, on which 
$\limsup_n|I_n(A,{\bf f,t},\ell)|\leq1/k$. Since the number of sets $A$ in (\ref{y1}) is finite, given $\ell$,
and since the set of sequences ${\bf t}$, with $t_i 
\in  \cup_\ell L^{-2\ell} {\N}$, is countable,
 $\CB (\bf f):=\cap_{\bf t} \cap_{A\neq\emptyset} \cap_k \CB_k({\bf f,t},  A)$   is a set 
of measure one on which
\rr
\lim_{m\to\infty}( \CE_{m+\ell}\prod_i f_i(
\omega (t_i))-E_\ell^{T_{m}}\prod_i f_i(L^{-\ell}\omega(L^{2\ell}
\omega (t_i)))=0
 \label{y7}
\qqq
where, as we recall from Section 2 (see (\ref{exps1})), 
$E_\ell^{T_m}$ is the expectation in the random walk
with transition probability $T_m$, in time $L^{2\ell}$; thus, the second term in (\ref{y7})
corresponds to the $A=\emptyset$ term in (\ref{y1}). 

We are left with proving a deterministic statement, namely 
that the second term in (\ref{y7}) converges to
$ \CE^D\prod_i f_i(
\omega (t_i))$.
Let again $t_i\in L^{-2\ell}\mathbb N$. Then,
\rr
 \CE^D\prod_i f_i(
\omega (t_i))=E_\ell^{T^*_D}\prod_i f_i(L^{-\ell}\omega(L^{-2\ell}
t_i)).
 \label{y8}
\qqq
Write
 $  T_m=T^*_D+\tau_m$. 
 Bounding the $f_i$'s as above, see  (\ref{y2}), we get that 
 the difference between the second term in (\ref{y7}) and (\ref{y8}) 
 is bounded by
 $$\sum_{k=1}^N( ^N_k)
\int dx\ ((\tilde T^*_D)^{N-k}\tilde \tau_m^k)(x)$$
where  $N=L^{2\ell}$ and the tilde is defined as above.
 By   (\ref{j4a}) and the explicit form  (\ref{j15}) of $T^*_D$, this sum is bounded by $C(\ell, L) \delta_m$
 and the claim follows.
 
\hfill $\Box$

\vspace*{4mm}

\no {\bf {Acknowledgments.}}
We thank  Carlangelo Liverani and Stefano Olla
for useful discussions. J. B. thanks the Belgian Internuniversity Attraction Poles Program  and A.K.  thanks the Academy of Finland
for financial support. We also thank the \'Ecole Normale Sup\'erieure (Paris), the University of Paris-Dauphine and the Institut Henri Poincar\'e (Paris), where part of this work was done, for their hospitality and financial support.

\end{document}